\documentstyle[11pt,newpasp,twoside,epsf]{article}
\def\edcomment#1{\iffalse\marginpar{\raggedright\sl#1\/}\else\relax\fi}
\reversemarginpar

\begin{document}
\title{The Evolution of the Interstellar Medium Around Young Stellar Clusters}
 \author{Daniela Calzetti}
\affil{Space Telescope Science Institute, 3700 San Martin Drive, Baltimore, 
MD 21218, USA, calzetti@stsci.edu}
\author{Christina A. Tremonti}
\affil{Dept. of Phys. \& Astron., The Johns Hopkins University, 3400 North 
Charles St., Baltimore, MD 21218, USA, and Space Telescope Science Institute, 
cat@pha.jhu.edu}
\author{Timothy M. Heckman}
\affil{Dept. of Phys. \& Astron., The Johns Hopkins University, 3400 North 
Charles St., Baltimore, MD 21218, USA, heckman@pha.jhu.edu}
\author{Claus Leitherer}
\affil{Space Telescope Science Institute, 3700 San Martin Drive, Baltimore, 
MD 21218, USA, leitherer@stsci.edu}

\begin{abstract}
The interplay between the ISM and the massive stars formed in clusters
and, more generally, in recent events of star formation is reviewed
via the global effects each has on the other. The pre-existing
environment affects the properties of the massive stars, the duration
of the star-forming event and could potentially affect the IMF. The
collective effect of massive-star winds and supernova explosions
creates a structured ISM by forming bubbles, supershells and, in more
extreme cases, by inducing large-scale gas outflows. Gas/dust removal
may quench star formation in young stellar clusters. Conversely,
supernova-driven shocks may trigger star formation in molecular clouds
surrounding the stellar clusters. Metal ejection from the massive
stars is responsible for the pollution of the ISM and, if the
metal-rich gas can escape the galaxy's gravitational potential, of the
IGM. The environment where stellar clusters form is populated by a
diffuse stellar population which contributes between 50\% and 80\% of
the total UV light. The investigation of the nature of the diffuse UV
light is the subject of a study employing HST STIS spectroscopy, whose
preliminary results are presented and briefly discussed.
\end{abstract}


\section{Introduction}

Stellar clusters are a prominent product of events of star
formation.  The study of young stellar clusters has recently received
new impetus after the suggestion that the most massive among such
systems may be progenitors of globular clusters (Kennicutt \& Chu
1988, Lutz 1991, Holtzmann et al. 1992, Whitmore et al. 1993,
O'Connell, Gallagher \& Hunter 1994, Meurer et al. 1995, Ho \&
Filippenko 1996, to quote a few). In this review, we discuss the
relation between young stellar clusters and the surrounding
environment, both gaseous and stellar. The interplay between massive
stars and ISM, for instance, plays a crucial role in the evolution of
a burst of star formation. Young stellar clusters are thus 
discussed here in the context of spatially extended, active
star-formation events (starbursts). Because of the vastness of the
topic, this review is necessarily incomplete, and many significant
contributions may have been, unwillingly, overlooked.

\section{The Effects of the Environment on the Stellar Clusters}

The environment affects or can affect to various degrees
characteristics of the stellar clusters/star-forming regions. 
Three such effects are briefly discussed below.

$\bullet$ The metallicity of the cloud from which the stars form
correlates with the massive stars' wind properties and with the number
of Wolf-Rayet stars produced. Both effects contribute to increase the
mass loss from the stars and the momentum and energy transfer into the
surrounding medium for increasing metallicity (Leitherer 1993). If the
metallicity is above solar, stellar winds are more important than
supernovae for the energy deposition rates in star formation episodes
younger than $\sim$20~Myr. From an observational point of view, the
dependence of the evolutionary history of the OB population on
metallicity should result in a correlation between the wind lines
produced by O-stars, such as SiIV($\lambda$1400~\AA) and
CIV($\lambda$1550~\AA), and the metallicity of the region. A
correlation between the strength of the combined equivalent widths of
SiIV and CIV and the oxygen abundance has been indeed observed in
starburst galaxies, as shown in Figure~1 (from Heckman et al. 1998,
see also Storchi-Bergmann, Calzetti \& Kinney 1994).

\begin{figure}
\plotfiddle{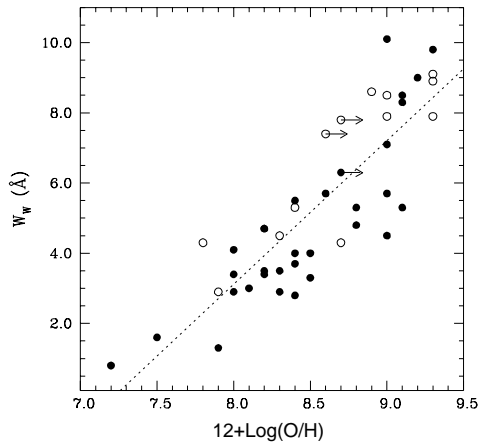}{2in}{0}{55}{55}{-130}{-115}
\caption{The sum of the SiIV($\lambda$1400~\AA) and
CIV($\lambda$1550~\AA) equivalent widths, W$_W$, as a function of the
oxygen abundance for a sample of starburst galaxies, from Heckman et
al. (1998). The two lines have been measured from IUE spectra of the
galaxies, and trace primarily the hot-star winds in systems with
active star formation, although a significant contribution from
interstellar absorption lines may also be present. The poor spectral
resolution of IUE does not permit, in fact, the discrimination 
between wind lines and interstellar lines.}
\end{figure}

$\bullet$ The gas supply determines the maximum duration of a
star-forming event.  This effect is not generally observable in
individual stellar clusters, which can be considered single-age
populations (with spreads $\le$10~Myr, e.g. Massey, Johnson \&
DeGioia-Eastwood 1995a). It is observable within star-forming regions
taken as a whole, like Giant HII regions (e.g., 30~Doradus) or
starbursts (e.g., NGC5253), where multiple stellar clusters and OB
associations spanning a range of ages are usually present
(e.g. Calzetti et al. 1997, Grebel \& Chu 2000). In rigorous
terms, the gas supply is not the only parameter to be considered for
the starburst duration, as more complex interactions between massive
stars and their surrounding environment take place (see next section).

$\bullet$ The relation between environment and stellar Initial Mass
Function (IMF) is probably the most controversial topic. A complete
review of the current status of IMF studies is contained in the
Proceedings of the 38th Herstmonceux Conference (Gilmore \& Howell
1998). The underlying reason for the ongoing controversy is that no
unique theory for the observed mass spectra exists yet (e.g. Clarke
1998), and observations are still not entirely effective at
constraining theories. If we parametrize the IMF as
$\phi$(m)$\propto$m$^{-\alpha}$, the slope in stellar clusters is
$\alpha\approx$2.3--2.5 for masses above M$\approx$1~M$_{\odot}$ (see
summary review by Kroupa, these Proceedings). The observed slope is
for all practical purposes a `Salpeter slope'. There appear to be no
systematic variations in the IMF slope between the Milky Way, the LMC
and the SMC stellar clusters, for masses above 5--10~M$_{\odot}$
(Massey et al. 1995a,b), nor a measurable variation in upper mass
cut-off, which is $>$70--100~M$_{\odot}$; this lack of a systematic
trend covers about a factor 10 in metallicity and a factor $\sim$1000
in stellar density (Massey 1998). However, the scatter around the mean
slope is $\Delta\alpha\sim$0.5, which is variously interpreted as an
effect of observational uncertainties, environment-dependent
variations (Scalo 1998), or stochastic scatter from a universal slope
(Elmegreen 1999). Below M$\approx$1~M$_{\odot}$, the IMF can be
described either as a log-normal distribution (Paresce \& De Marchi
1999) or by two power-laws, with much flatter slopes than Salpeter,
$\alpha\approx$1.5 for M$<$0.5~M$_{\odot}$ and $\alpha\sim$2.2 for
0.5~M$_{\odot}<$M$<$1~M$_{\odot}$ (e.g., Kroupa, Tout \& Gilmore
1993), in both cases with a smaller scatter than in the higher mass
region.

\section{The Effects of the Stellar Clusters on the Environment.}

The impact of stellar clusters on the surrouding ISM is mediated 
by the effects of hot-star winds and supernova explosions in terms of 
mass and metal deposition, and energy and momentum transfer. 

$\bullet$ In an evolving burst of star formation, the energy injected
into the ISM by the collective effect of massive-star winds and SNa
explosions may sweep up material and produce structures in the form of
bubbles, supershells, and filaments (Shull 1993, Chu \& Kennicutt
1994). In more extreme cases, large-scale outflows of material, or
superwinds, may be produced (Heckman, Armus \& Miley 1990). 30~Doradus
in the LMC, which has been called the `Rosetta Stone of Starbursts'
(Walborn 1991), is the prototype of turbulent environment; here the
filamentary interfaces between `evacuated' regions and surrounding
dust clouds dominate the nebular emission (Scowen et al. 1998, Walborn
et al. 1999). The gas kinematics reveal the complex structure of the
30~Dor region: slow expanding shells (v$<$100~km~s$^{-1}$ and sizes up
to $\sim$100~pc), driven by stellar winds and SNe, are present
together with fast expanding shells (v$\sim$100--300~km~s$^{-1}$ and
sizes up to $\sim$30~pc), these most likely driven by SNRs (Chu \&
Kennicutt 1994). The massive-stars- and SNe-induced gas removal from
the stellar cluster's site will have the effect of quenching
star formation in that region, thus producing a large-scale
self-regulating mechanism for the star formation intensity (e.g.,
Kennicutt 1989, Heckman 1997).

$\bullet$ The energy and momentum injection by evolving and exploding
massive stars can also cause destruction/removal of dust from the
site of star formation. While gas outflows can eject significant
amounts of both interstellar gas and dust from the star-forming
region, shocks from SNe can be responsible for the dust grain
destruction, via grain-grain collisions and sputtering (Jones et al.
1994). As a net result, the main source of opacity for the
starburst and the stellar clusters is the dust {\em surrounding} the
region of star formation, rather than the dust inside the region.  The
dust geometry will then be equivalent to a shell surrounding a central
light source (Calzetti, Kinney \& Storchi-Bergmann 1996).

$\bullet$ Hot-star winds and SNa explosions can drive ionization
and shock-fronts through molecular clouds, causing star formation to
propagate (Elmegreen \& Lada 1977). Although controversy exists on the
theory of star formation propagation (McCray \& Kafatos 1987),
triggered star formation has been advocated for the multiple
generations of stellar clusters observed in 30~Doradus (Hyland et
al. 1992, Walborn et al. 1999, see also Grebel \& Chu 2000) and in starburst
galaxies like M82 and M83 (Satyapal et al. 1997, Puxley, Doyon, \&
Ward 1997), where time scales for star formation have been measured to
span a few tens of Myr.

$\bullet$ The material ejected by supernovae pollutes the ISM with
about 1~M$_{\odot}$ of metals for every $\sim$40~M$_{\odot}$ of new
stars formed, assuming a Salpeter IMF. Effects of localized metal
pollution due to the presence of young stellar clusters have been
observed by Kobulnicky et al. (1997) in the nearby starburst dwarf
NGC5253, where an enhancement of Nitrogen over Oxygen by a factor 3
has been measured in regions in the immediate proximity of a stellar
cluster. If gas outflows from the region of star formation develop
into large-scale superwinds, these can carry metals out of the galaxy
and contribute to the enrichment of the IGM. This mechanism will be
more efficient in low-mass galaxies, where the escape velocity is
smaller than in more massive galaxies. A starburst with a mechanical
luminosity of 10$^{38}$~erg~s$^{-1}$, hosted in a 10$^9$~M$_{\odot}$
galaxy, will produce enough energy to eject about 70\% of the newly
formed metals into the IGM (MacLow \& Ferrara 1999); this fraction
obviously increases for more luminous starbursts and for decreasing
galaxy mass.

\section{A Case Study: Stellar Clusters in NGC5253}

In starburst regions, young stellar clusters are generally embedded
within diffuse UV light, which represents about 50\%, and up to 80\%,
of the UV brightness of the region (Conti \& Vacca 1994, Hunter et
al. 1994, O'Connell et al. 1994, Meurer et al. 1995). The nature of
the diffuse UV light light is still unclear; scattered light from the
stellar clusters, small unresolved stellar clusters, a diffuse stellar
population originating as an independent mode of star formation, or
stars left behind by dissolved clusters are all possible
intepretations.

\begin{figure}[t]
\plotfiddle{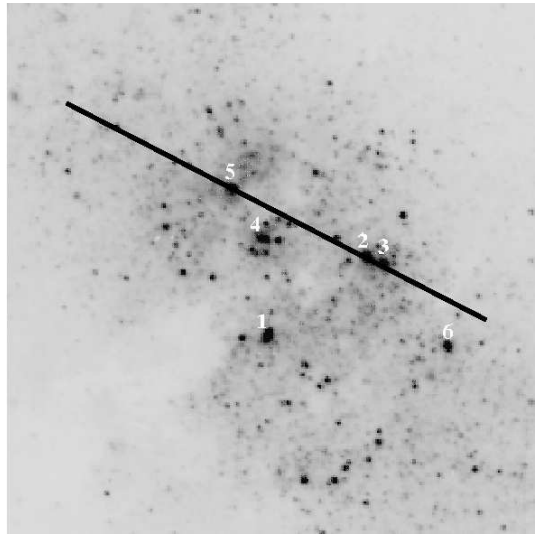}{2.5in}{0}{40}{40}{-120}{-60}
\caption{HST WFPC2 image of the central $\sim$500~pc of NGC5253, in
the F547M filter ($\sim$V). North is up, East is left. The numbers
label the six brightest stellar clusters, whose general properties are
discussed in Calzetti et al. (1997). The lighter region crossing the
center of the image from SE to NW locates the position of the dust
lane. The black line marks the approximate position of the STIS
0.1$^{\prime\prime}$ slit used to obtain UV spectra of clusters~2, 3,
and 5, and of the field surrounding the clusters, with an approximate
spectral resolution of $\sim$2~\AA.}
\end{figure}

Long-slit ultraviolet spectra of the central starburst in the nearby
(D$\sim$4~Mpc) dwarf galaxy NGC5253, obtained with STIS on-board HST,
are being employed to investigate the nature of the diffuse UV light
(Tremonti et al. 2000). NGC5253 is a ``benchmark starburst'', with
active star formation concentrated in the central
$\sim$20$^{\prime\prime}$ (about 400~pc at the distance of the
galaxy), superimposed on an older, quiescent stellar population.  The
central star-forming region is very blue, although it is crossed by
dust lanes which produce patchy and heavy obscuration.  The starburst
region contains about a dozen bright stellar clusters (Meurer et
al. 1995), the six brightest of which are identified in Figure~2 (from
Calzetti et al. 1997). Clusters 4 and 5 are located within the most
active part of the starburst, which is $\sim$100~pc in size and has a
star formation rate density of
$\sim$10$^{-4}$~M$_{\odot}$~yr$^{-1}$~pc$^{-2}$, corresponding to the
maximum levels observed in starburst galaxies (Meurer et
al. 1997). The H$\alpha$ emission from the galaxy has almost perfect
circular symmetry around cluster~5, which is thus driving the
ionization. This cluster is deeply buried in the dust lane crossing
the center of the galaxy, with an optical depth A$_V\approx$30~mag
(Calzetti et al. 1997).

Previous HST WFPC2 broad- and narrow-band imaging of the center of
NGC5253 at both optical and ultraviolet wavelengths had yielded
tentative ages for the six stellar clusters of Figure~2. In
particular, the colors and the H$\alpha$ emission of cluster~5
suggested this cluster to be the youngest among the six, with an age
$<$3~Myr; clusters~2 and 3 came out to be the oldest, with ages
estimated between 30 and 60~Myr (Calzetti et al. 1997). These three
clusters, together with the field between them, are the targets of 
our STIS UV spectroscopic observations, as shown in Figure~2.

\begin{figure}[t]
\plotfiddle{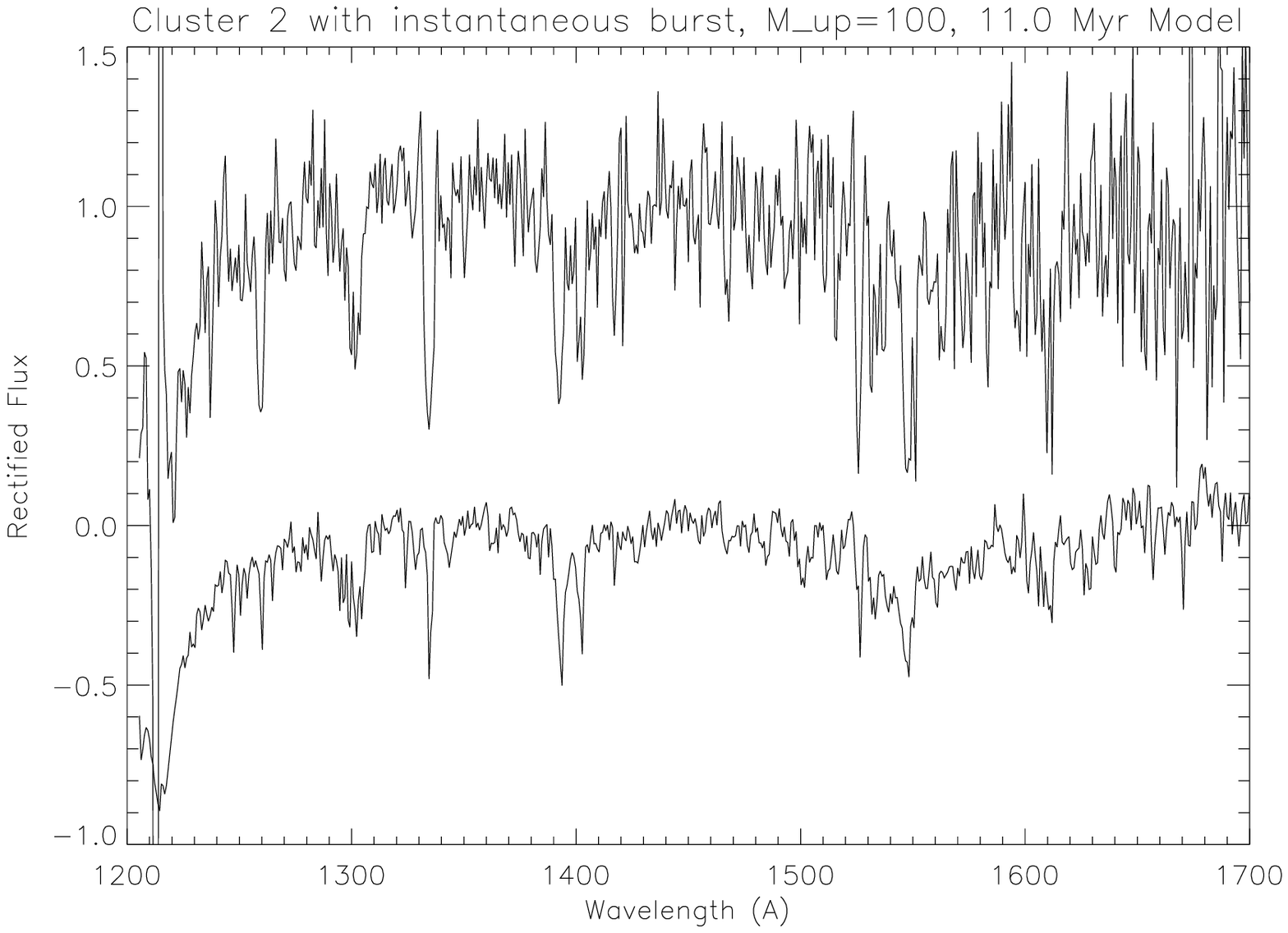}{1.5in}{0}{40}{40}{-250}{-170}
\plotfiddle{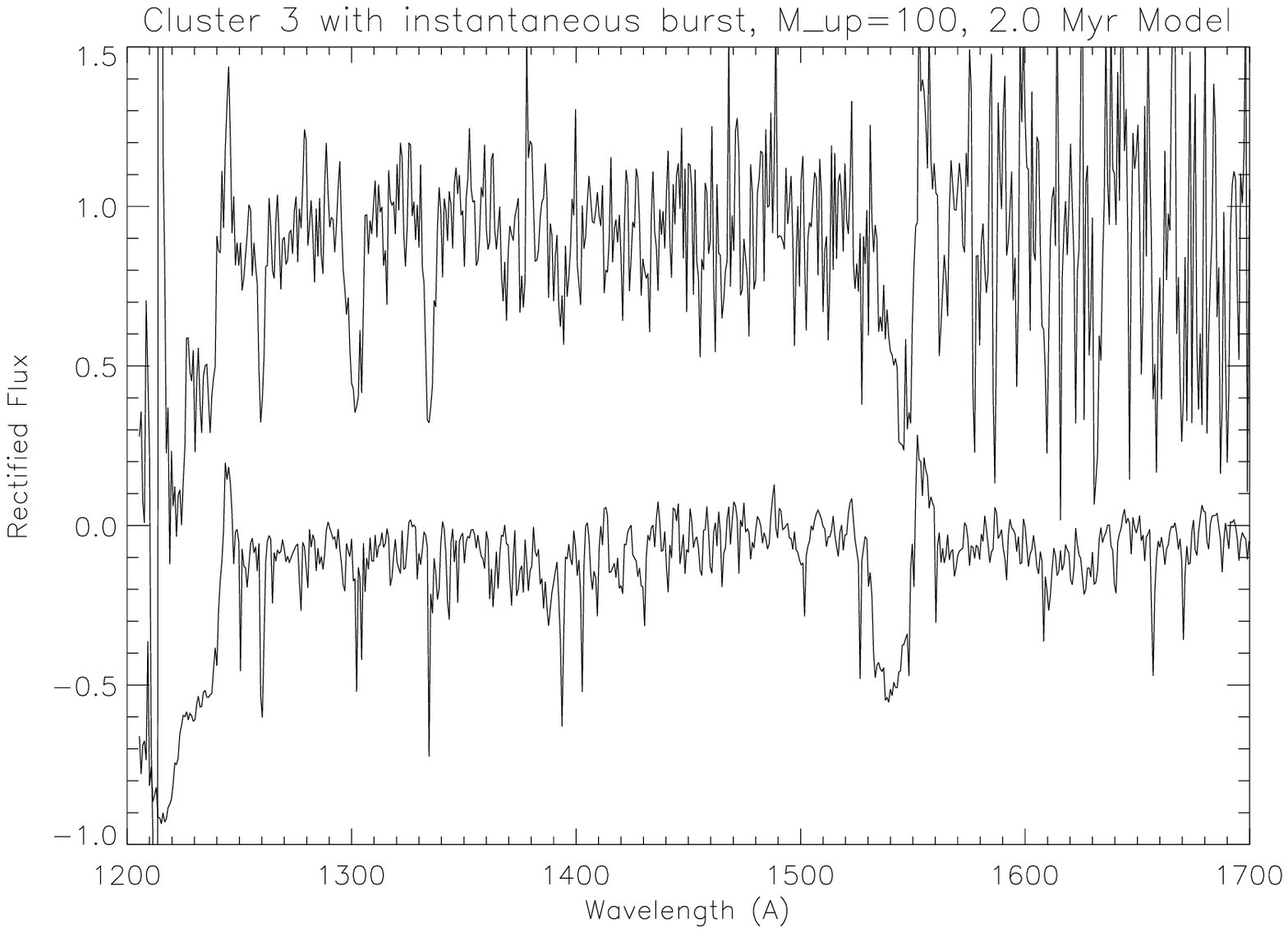}{0in}{0}{40}{40}{-40}{-147}
\caption{Rectified UV spectra of clusters~2 (left) and 3 (right). In
each panel, the data (top) and the preliminary best fit stellar
population models (bottom) are shifted relative to each other.  The
best fit instantaneous-burst stellar populations have ages 11~Myr and
2~Myr for cluster~2 and 3, respectively, using a Salpeter IMF with an
upper mass cut-off of 100~M$_{\odot}$.}
\end{figure}

The rectified UV spectra, in the wavelength range 1,200--1,700~\AA, of
clusters~2, 3, and 5, and of the field between them are shown in
Figures~3 and 4. Preliminary fits to the spectra using the stellar
population synthesis models of Leitherer et al. (1999, recently
updated to include B stars), are shown in the same figures,
down-shifted relative to the observational data.  Instantaneous burst
populations have been adopted for all the spectra, although, for the
field, the case of continuous star formation is currently being
investigated. The metallicity used in the fits is 0.2~Z$_{\odot}$, a
good match to the metallicity of the gas in the galaxy, which is about
1/6~solar. A range of IMF slopes and upper mass cut-offs are also
being tested on the data; the preliminary results reported here use a
Salpeter slope and upper mass cut-off of 100~M$_{\odot}$.

\begin{figure}[t]
\plotfiddle{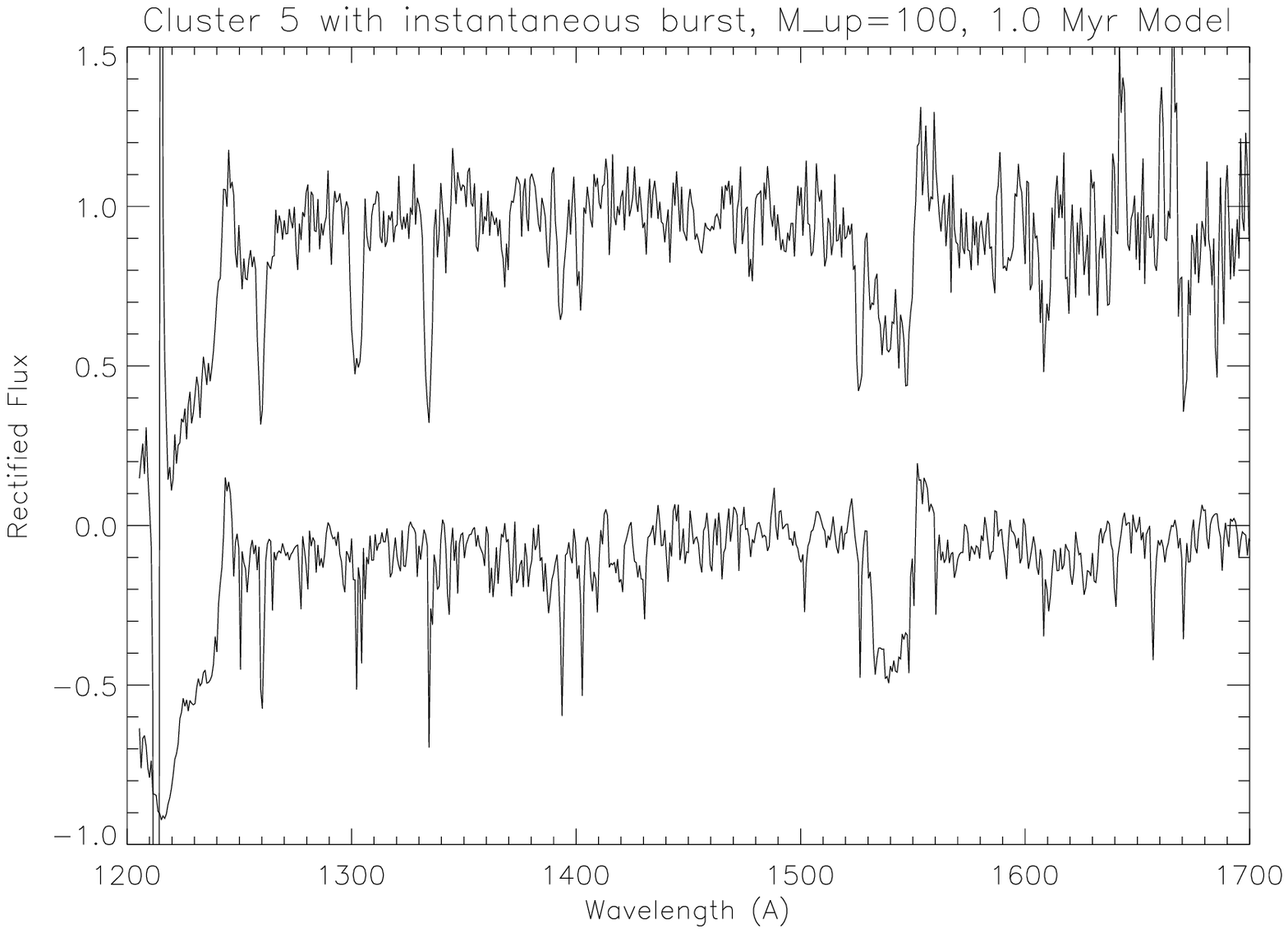}{1.5in}{0}{40}{40}{-250}{-170}
\plotfiddle{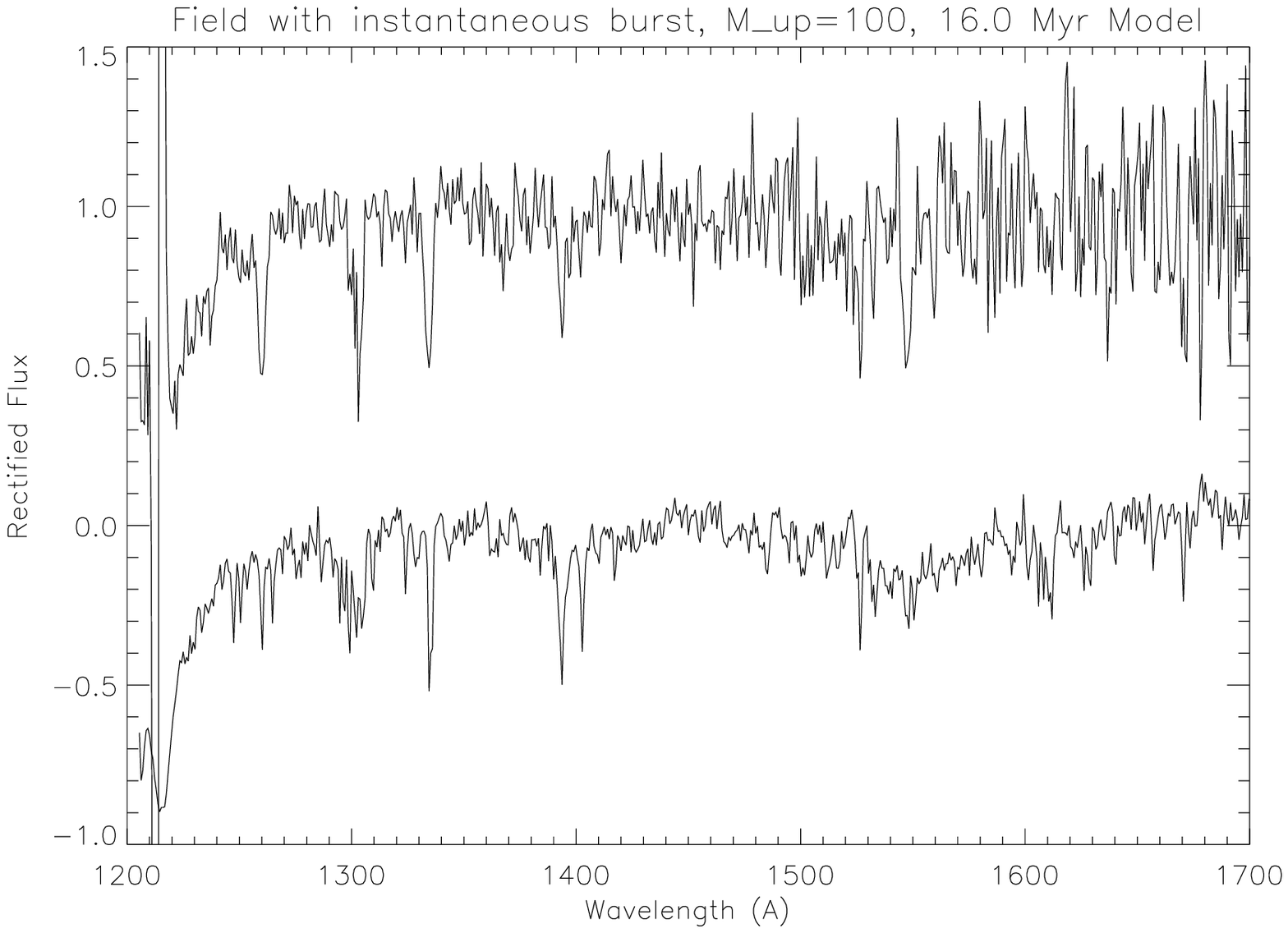}{0in}{0}{40}{40}{-40}{-147}
\caption{As in Figure~3, for cluster~5 and for the field surrounding
the three clusters. The best fit instantaneous-burst stellar
population gives an age $\sim$1~Myr for cluster~5 and 16~Myr for the field.}
\end{figure}

The best fit ages of clusters~2 and 3 are 11~Myr and 2~Myr,
respectively, a factor 5 and 20 younger than the ages inferred from the
HST broad-band colors. For instance, the presence of P-Cygni profiles
in the lines of NV($\lambda$1240~\AA) and CIV($\lambda$1550~\AA)
exclude ages older than $\sim$6~Myr for cluster~3. Cluster~5 is
confirmed to be very young: the best fit model to the UV spectrum gives
an age $\sim$1~Myr. The very red spectral continuum also confirms that the
cluster is reddened by a considerable amount of dust; depending on the
dust obscuration correction, the inferred mass for cluster~5 is
between a few times 10$^5$ and 10$^6$~M$_{\odot}$; this cluster is
thus a `bona fide' super-star-cluster candidate (Calzetti et
al. 1997).

The first remarkable property of the field's UV spectrum is that the
lines corresponding to NV, SiIV and CIV are different, both
qualitatively and quantitatively, from the stellar lines in the
clusters' spectra. This excludes the possibility that the field UV
light is dominated by scattered cluster light. This is further
supported by the fact that the HST WFPC2 UV image shows resolved small
clusters and bright isolated stars in the region we call the
`field'. A preliminary best fit model using an instantaneous burst
population with Salpeter IMF gives an age of 16~Myr. The stellar
population in the field, thus, appears older than the populations in
the bright stellar clusters. Cluster~4 has, indeed, an age
$\sim$4~Myr, and clusters~1 and 6 have ages $<$15--17~Myr (Calzetti et
al. 1997). The old age of the field may indicate that its stellar
population originates from unresolved stellar clusters older than the
6 brightest ones, or from stars left behind by dissolved clusters. An
alternative interpretation is that the IMF of the field is drastically
different from the IMF of the stellar clusters (e.g., Massey et
al. 1995b). All three possibilities are currently under investigation.

\section{Summary}

The investigation of the interplay between recently formed stellar
clusters and the surrounding ISM is the key to understand
the evolution of star-forming events. For instance, the duration of a
burst of star formation is determined not only by the gas supply, but
also by the feedback of the massive-star winds and SNa 
explosions into the ISM, which may quench or possibly trigger new star
formation. In this review, we have highlighted, among other things,
the impact of the pre-existing environment on the properties of the
newly formed massive stars and presented some of the ongoing
controversy on environment-dependent IMFs. The feedback of stellar
clusters onto the surrounding ISM can be summarized as the collective
mass, metals, energy and momentum deposition by massive stars; the
general effects will be the creation of a highly structured
environment and the metal pollution of both ISM and IGM.

Diffuse UV-bright stars are generally found in the same environment
where young stellar clusters are located. Understanding the nature of
the diffuse population is key to understanding star formation
mechanisms; competing theories include unresolved stellar clusters, 
dissolved stellar clusters, and a second mechanism for star formation. We 
have briefly presented the ongoing analysis of recently obtained HST 
STIS UV spectra of stellar clusters and diffuse population in the nearby 
starburst galaxy NGC5253; the goal of the analysis is to constrain the above 
scenarios. 

{\bf Aknowledgements:} Daniela Calzetti thanks the Conference
Organizing Committee, and especially Ariane Lan\c{c}on, for the
invitation to this very stimulating meeting. The STScI Director's
Discretionary Research Funds are acknowledged for supporting the trip
and this work.

\vspace{0.3cm}

{\bf Q. (Daniel Schaerer):} From optical spectroscopy we know that your 
knots \# 4 and 5 in NGC5253 have some WR stars, which have allowed us to 
determine ages of $\sim$2--5~Myr. Could you please comment on the difference 
with the very young age you determine for cluster \# 5? Also, it would be 
interesting to get an optical spectrum of cluster \# 3, for which you find 
an age young enough to expect WR stars. Do you have such data?

{\bf A.:} The age we derive for cluster \# 4 from HST broad-band
colors and H$\alpha$ emission is $\sim$2.5--4.4~Myr, compatible with
the presence of WR stars. The position of your slit misses the center
of cluster \# 5, so our age determination for this cluster is not
necessarily incompatible with your detected WR stars. Cluster \# 5
`sits' on a region of very active star formation, and I believe not
unlikely that multiple generations of stars are contemporary present
in that region. We don't have an optical spectrum of cluster \# 3; it
would be interesting to get one, in order to address your last
question.
\end{document}